\documentclass{appolb}
\usepackage{url}
\usepackage{algorithm}
\usepackage[usenames, dvipsnames]{color}
\usepackage{multicol}
\usepackage{algorithmic, amsmath}
\usepackage{epsfig,graphicx}
\usepackage{enumerate}
\usepackage{adjustbox}
\usepackage{calligra}
\usepackage{mathptmx}       
\usepackage{helvet}         
\usepackage{courier}        
\usepackage{type1cm}        
\usepackage{makeidx}         
\usepackage{graphicx}        
\usepackage{multicol}        
\usepackage[bottom]{footmisc}

\usepackage[figuresright]{rotating}

\begin{document}
\title{Effect of correlations on routing and modeling of Time Varying Communication Networks}
\author{Suchi Kumari, Anurag Singh
\address{Department of Computer Science and Engineering,\\
National Institute of Technology,\\
Delhi, India\\ email: {suchisingh@nitdelhi.ac.in, Corresponding author: anuragsg@nitdelhi.ac.in}}
}
\maketitle
\begin{abstract}
Most of the real world networks are complex as well as evolving. Therefore, it is important to study the effect of network topology on the dynamics of traffic and congestion in the network. To account this problem, we have designed a time-varying network model where a new node will join a node in the existing network with probability proportional to its degree and disassortativity with its neighbors. Betweenness centrality (BC) plays an important role to find the influential node and user's shortest route in the network. As shortest route comprised of hub nodes and chances of congestion is more on these nodes. Hence,  BC-BC correlation is used to find user's route. A connection between two hub nodes reduces the data forwarding capacity of connecting link with higher probability. If a node shows disassortativity with its neighbors then it may forward more packets and may be chosen for routing. Furthermore, user's optimal data sending rate as well as critical packet generation rate of the proposed model is calculated and shown improved results in comparison than the classical scale-free network model.
\end{abstract}
   
\section{Introduction}

The structure of the complex systems can be analyzed by using the graph based methodologies, where nodes represent elements of the system and links denote the interactions among the nodes. A model to fulfill the same is developed by Barabasi-Albert \cite{bara}. Study the dynamics of such networks has received a great wave of interest for the researchers in past few years. It is a challenging task to design a congestion free time varying communication networks (TVCN) to handle dynamic traffic. Therefore, the main aim of this paper is to design a well structured (with minimized congestion) TVCN model and provide an efficient route to each user in the network. Various network properties such as degree distribution, clustering coefficient, diameter are used to represent the physical structure of the network. In order to analyze the designed network model, these properties should be measured and compared with the existing real world systems.  Like all the other discussed properties, Degree degree correlation (DDC) is one of the important statistical properties which describes the topology of scale free (SF) networks. This quantity measures the tendencies of nodes to connect with other nodes that have similar (or opposite) degrees as themselves. TVCN is designed with the consideration of both DDC and preferential attachment. Betweenness Centrality (BC) plays a very important role in finding user's shortest route hence, like DDC, BC-BC correlation may be used to assign a shortest as well as efficient route to each user. 


Today, the Internet has changed the way of communication, business, studies and many other activities. At the early stage, the Internet was quite small and its structure was simple but, it evolves with time and it is very difficult to communicate data efficiently in this scenario. Erdos-Renyi model is the first graph based model where a node will become highest degree node in the network is entirely by chance \cite{ER}. Barabasi and Albert offers a more realistic model of the network and has given the concept of evolving network with preferential attachment and follows power law degree distribution, called as scale free (SF) networks. Evolving nature of the graph can be shown by attaching time parameter with the network representation, $ G(N, E, t) $. Many time-varying graphs (TVG) are generated by using the model proposed \cite{TVG,TVGdynamic}. Wehmuth \textit{et al.} \cite{TVG} proposed a unifying model for representing finite discrete TVGs. Kohar \textit{et al.}\cite{synchronization} studied the stability of the synchronized state in time-varying complex networks and found that the time taken to reach synchronization is lowered and the stability range of the synchronized state increases. A framework is designed to obtain degree distribution of evolving network with the consideration of deletion of nodes \cite{2007continuum}. In \cite{2015partial}, a new node has access to a fraction of nodes and a new connection is formed with a probability proportional to the degree of the nodes. A framework to represent mobile networks dynamically in a Spatio-temporal fashion is designed and algebraic structural representation is also provided by \cite{kumari2016}. A TVCN is designed where nodes and links are getting added into the networks while restructuring is performed in the existing networks through rewiring and removal of the links \cite{kumari2017}. 

SF model is the accepted theory of the evolution of networks but in real world systems, the theory does not match with the data. The SF model has been used to analyze a wide variety of graphs but on close examination, a number of theoretical drawbacks are found. SF model implies that the degree distribution follows the power law with exponent, $ \gamma = 3 $ but in many real world networks the exponent, $ \gamma $ has values between $ 1.2 $ and $ 2.9 $ \cite{tee2016}. BA model shows that older node always has the most links and a later node can never become the hub node in the network. But in a real scenario, the growth rate of a node does not depend only on its incoming time but also on some other parameters such as its own intrinsic properties that affect the rate at which a node attain links. A model is developed by Barabasi in collaboration with Bianconi where evolving networks is mapped into an equilibrium Bose gas. In this model \cite{BB}, each new node is assigned a random fitness parameter, $ \eta_i $ and it connects to a node $ i $ based on the product value of node $ i$'s degree, $ k_i $ and its fitness, $ \eta_i $. The dependency on $ \eta_i $ implies that between two nodes with the same degree, the one with higher fitness is selected with a higher probability and a younger node can also acquire links rapidly if its fitness value is higher than others. The nodes with the highest fitness turn into the largest hubs in the network with time. The node's ability to acquire links affect the topology of the network \cite{networkscience}. BA model assumes that a graph will evolve indefinitely without considering any constraint or limit on it.  

Apart from degree distribution, degree-degree correlation (DDC) is a network property in which nodes with similar attributes, such as degree, tend to be connected. The DDC has an important influence on the structural properties of the network and hence, used to measure stability, robustness \cite{multirobust}, controllability of the network \cite{2016analysis,multiplex}, spreading of diseases \cite{Holme2016}, the traffic dynamics on networks and other time varying processes. DDC is used to divide SF networks into three types: assortative, disassortative and neutral networks \cite{newmanassortative}. For assortative networks, hubs(small degree nodes) tend to link to other hubs (small degree nodes) and avoid small degree nodes (hubs). In a disassortative network, hubs (small degree nodes) avoid each other, linking instead to small-degree nodes (hubs). While in the neutral network, number of links between nodes are random. Social network such as actor, email, mobile phone, science collaboration network etc, and citation network are examples of an assortative network. The communication network such as the Internet and World Wide Web, Biological network such as protein interaction, metabolic network etc. are considered as a disassortative network. Power grid network is considered as a neutral network.    
 
Networks such as social, biological systems consist of multiple layers of networks interacting with each other and is known as multiplex networks where a node belongs to multiple layers with different types of links. Network layers are correlated with each other and a positive correlation indicates that a hub in one layer also has many neighbors in the other layers. The role of correlations to understand various robustness properties of multilayer network is studied in \cite{multirobust}. Positively correlated multiplex networks are more robust than negatively correlated with random failure whereas negativity correlated network is more robust for the targeted attack. There is a set of influential nodes that control the whole network and is called a dominating set. To control the whole network the size of the dominating sets should be minimum. DDC is used to calculate the size of a minimum dominating set (MDS) to measure network controllability \cite{2016analysis}. Negative correlation reduces the size of an MDS and enhances network controllability, whereas positive correlations hardly affect the size of an MDS. Furthermore, apart from the controllability issue, the developed techniques provide new ways of analyzing complex networks with DDC.  In \cite{multiplex}, researchers explored that the multiplex networks with positive correlation are easier to control for the small density of interaction while dense interactions can't control the multiplex networks. Each node has different information processing capability and the node with highest processing capability is known as the most influential node in the network. The identification of such set of nodes is an important task to control the spreading processing on the network. Rewiring is done randomly in SF networks to analyze its topological properties as well as on the DDC of the network \cite{ranrewiring}. Random rewiring reduces average degree of nearest neighbors of high degree nodes and results to increase in disassortativity of the network.

Multiple users want to access resources in communication networks hence, fair sharing of resources is very important. Kelly \cite{kelly} and other researchers \cite{la,incentives} provided rate allocation schemes to assign resources in a fair manner. In this scheme, a utility function is associated with each user and it is considered as being fair if it maximizes the sum of utilities of all the users in the network. Evolving network model only considers the addition of nodes \cite{bara}, although some works consider deletion of nodes during/beyond the evolution of the networks \cite{chen2004,2007continuum} and \cite{kumari2017,kumari2017optimal,kumari2016modeling} considers an addition, rewiring, and removal of the links. All these approaches are an extension of BA model where addition and rewiring are done on the basis of preferential attachment while deletion considers anti-preferential attachment. In Real life networks like the Internet, communication networks, WWW, transportation networks, new links appear due to the addition of a new node and few links are removed due to break down and to reduce maintenance cost while few links are rewired to handle dynamic traffic. In this context, the proposed work considers designing of DTVCN model to minimize network congestion with maximization of network utility. As betweenness centrality (BC) is one important parameter to find out congestion at a node in the network hence, we have considered BC-BC correlation for routing. User's data rate is maximized by selecting the shortest path whose nodes are the least BC-BC correlated with its neighbors. Least correlation with neighbors reduces overall congestion at the node and the path will give maximum optimal rate to that user. 

Section 2 describes the methods used for measuring correlation, provides a traffic model and provides a classical mathematical model used in the analysis of rate control behavior of the user's route. Section 3 introduces a model for growth dynamics of DTVCN, formulation of scaling exponent and the proposed routing strategies for the communication network. Section 4 presents theoretical and simulation results, and in Section 5, conclusions and future research plan are discussed.

\section{Background and Related Work}
The communication network is assumed as time varying in this paper. Hence, a brief description is presented in this section about mathematical representation of the TVCN. Static network, $ G(N,E) $ can be represented by using only two parameters, nodes ($ N $) and links ($ E $). While, in the case of time varying network, $ G(N,E,T) $ we need to add one more parameter i.e, time. The set of nodes is represented by $ N $, $ E $ represents the total number of links and lifespan of the network is $ T $. A set of $ R $ users is willing to access and send data at this time varying communication network (TVCN). Each user wants to send some data from the source node, $ s $ to the destination node, $ d $. In TVCN, a route is assigned to each user $ r \in R $ for a time instant $t_i \in T $. At the end of $ t_i^{th} $ time, a zero-one matrix $ A $ of the size $( N \times N )_{t_i} $ is defined where, $ A_{i, j} (t_i) = 1 $, if nodes $ i $ and $ j $ are connected at time $ t_i $ otherwise zero. A link, $ e_{mn} $ between node $ m $ and node $ n $ that appears in the user's route can send maximum $ C_{e_{mn}} $ units of data through it, where, $ C_{e_{mn}} $ is the capacity of link $ e_{mn} $. The congestion and user's path are selected based on the correlation parameter between nodes in the network. 
Here, the correlation between the nodes is considered as degree degree correlation (DDC).

\subsection{Methods for Measuring Degree Degree Correlation}
Correlation is defined using covariance and both are used to show how two random variables are related to each other. Covariance indicates whether two variables are directly related or inversely related while, correlation returns the degree to which variables tend to move together. Correlation between two random variables $ X $ and $ Y $ is expressed as\\
$ Cor(X,Y) = \frac{Cov(X,Y)}{\sqrt{Cov(X,X) \times Cov(Y,Y)}} $ where, $ Cov(X,Y) = \frac{\sum_{i=1}^{n} (X_i-\bar{X})(Y_i-\bar{Y})}{n-1} $. \\
Here, in this paper, these variables are considered as degrees of nodes.
\begin{enumerate}[1.]
\item One of the widely used measures for degree degree correlation is average neighbors degree of a node (ANDN) in the network. Let us define a conditional probability $ P_c(k'|k) $ such that a link having one end node with degree $ k $ will point to a node with degree $ k' $. A direct measurement of $ P_c(k'|k) $ is difficult due to large fluctuations in the value of the degrees of nodes hence, it is used to extract some other features i.e, average degree of neighbors of the nodes with degree $ k $, represented by, $ \langle k_{nn} \rangle = \sum_{k'} k' P_c(k'|k) $. There is a power law dependence between connectivity and degree, $ \langle k_{nn} \rangle \sim k^{-\nu} $ with $ \nu \simeq 0.5 $ \cite{dynamicalcor}. DDC is measured in terms of mean degree of neighbors of degree $ k $ and written as $ \langle k_{nn} \rangle (k) $. For assortative (or disassortative) network,  $ \langle k_{nn} \rangle (k)$ increases (or decreases) with increase in $ k $ while in uncorrelated network $ \langle k_{nn} \rangle (k)$ is independent of $ k $ \cite{dynamicalcor}. This representation provides only a qualitative characterization for the DDC in networks. Pearson provided a quantitative value of correlation coefficient, $ r_{deg} $ with the help of a correlation function \cite{newmanassortative}.

\item Let $ p_k $ is the probability of a randomly chosen node having degree $ k $. If a random link is chosen then the probability of this link to incident on a node, $ i $ is proportional to its degree $ k $ and also depends on degree distribution, $ p_k $, and this probability is represented by, $ \frac{k p_k}{\sum_k kp_k} $. Similarly, average degree of a neighbor is written as $ \sum_k k \frac{k p_k}{\sum_k kp_k} $. After reaching the node $ i $, degree distribution of the reached node is calculated and is called as excess degree distribution. If degree of a node, $ i $ is $ k $ then, total degree will be $ k+1 $ and excess degree distribution is $ q_k = \frac{(k+1) p_{k+1}}{\langle k \rangle} $. Excess degree distribution, $ q_k $ plays very important role in finding giant component of the network. Here, $ q_k $ is used to calculated DDC of a node and to find network's correlation coefficient, $ r_{deg} $. Let us find the joint degree distribution, $ p_{e_{jk}} $ of a randomly chosen link, $ e_{jk} $ of the two nodes at either end of the link. For uncorrelated network, the value of $ p_{e_{jk}} = q_j q_k $ but it will be different for assortative (or disassortative)  networks. The value of $ r_{deg} $ can be written in the form of correlation function as $ r_{deg} = \langle jk \rangle- \langle j \rangle \langle k \rangle = \sum_{jk} jk(p_{e_{jk}}-q_jq_k) $. There exists various kind of networks hence, the normalized value of correlation coefficient, $ r_{deg} $ is defined as \cite{newmanassortative},
\begin{equation}
r_{deg} = \frac{1}{\sigma (q)^2} \sum_{j,k} jk\{ p_{e_{j,k}}- q_jq_k \}. \label{e1}
\end{equation}
Here, $ \sigma (q)^2 = \sum_k k^2 q_k - {[\sum_k k q_k]}^2 \mbox{ and } -1 \leq r_{deg} \leq 1 $. The value of $ r_{deg} $ is positive (or negative) for assortative (or disassortative) network while zero for uncorrelated networks.
\end{enumerate}

\subsection{Network Traffic model}
A node, $ i $ may generate packets with rate $ \lambda_i $ and forward packets according to its capacity, $ C_i = \sum_{j \in Ne(i)} C_{ij} $ where, $ Ne(i) $ is the neighbors for node $ i $. The capacity of the network, $ C $ is calculated by summing up the capacity of each node. The sum of packet generation rate $ \lambda_i $ of each node $ i $ is termed as the load of the network, $ \lambda $ \cite{onset}. There are three possible relationships between $ \lambda $ and $ C $: (i.) $ \lambda < C $ implies that the system in free flow state, (ii.) $ \lambda = C $ shows the boundary case for congestion and (iii.) $ \lambda > C $ allows a system in congestion. If a node generate more packets than its capacity then that node will be congested and these nodes increase overall network congestion.  Zhao \textit{et al.} \cite{onset} proposed a model in which packet forwarding rate, $ C_i $ of a node $ i $ is calculated by introducing two models. First model considers degree of the node $ i $ and $ C_i $ formulated as, $ C_i = 1 + \lfloor \beta k_i \rfloor $. While, second model calculates $ C_i $ based on betweenness centrality and is given as, $ C_i = 1 + \lfloor \frac{\beta g(i)}{|N|} \rfloor $. Here, $ k_i $ = degree of node $ i $, $ g(i) $ = betweenness centrality of node $ i $, $ |N| $ size of the network and $ \beta $, $ 0 < \beta < 1 $. By theoretical estimation, the value of critical rate is given by $ \lambda_c = \frac{C_{Lmax} (|N|-1)}{g(Lmax)} $. The node with maximum packet forwarding capacity and betweenness centrality are denoted by $ C_{Lmax} $ and $ g(Lmax) $, respectively.\\
An order parameter $ \zeta(\lambda) $ \cite{rateeqn}, is used to describe the traffic and is given by,
\begin{equation*}
 \zeta(\lambda) =  lim_{t \rightarrow \infty} \frac{C}{\lambda}\frac{\langle \Delta Num_p  \rangle}{ \Delta t}
\end{equation*} 
$ \Delta Num_p = Num_p(t + \Delta t)-Num_p(t) $, $ Num_p(t) $ is number of packets at time $ t $ and $ \langle . \rangle $. This shows that average value over the time window $ \Delta t $. $ \langle \Delta Num_p  \rangle = 0 $ indicating that there is no packet in the network for $ \zeta = 0 $, system is in free flow state. 

Kelly \cite{kelly} formulated the rate allocation problem into optimization problem for a static network. We have updated the formulation of a classical model for the case of dynamic communication networks for the proposed DTVCN model.

\subsection{Updated classical mathematical model for rate allocation problems in TVCN}
The utility function, $U_r(x_r (t_i))$ for user $ r $ is an increasing and strictly concave function of $x_r (t_i)$ over the range $x_r (t_i) \geq 0$. Aggregate system utility is calculated by summing up all utilities of user $r$ and is denoted as $\sum_{r \in R} U_r(x_r (t_i))$. System utility can be optimized by using the following rate allocation problem.
\begin{eqnarray} 
&& SYSTEM(U(t_i),A(t_i)) \nonumber \\
&& maximize \sum_{r \in R, t_i} U_r(x_r (t_i)) \label{e2} \\
&& A^ Tx(t_i) \leq C(t_i) \mbox{ and }  x(t_i) \geq 0  \nonumber
\end{eqnarray}
$A(t_i)$ is a connection matrix at time $t_i$. The given constraint states that a link can not send data more than its capacity \cite{kelly}. It is difficult and unmanageable to allocate a suitable utility function and optimal rate to distinct users in complex networks. Hence, Kelly has divided this problem into two simpler problems named as user's optimal problem and network's optimal problem \cite{kelly}.  At each time instant if user wants to access a link $e_{mn} \in E$ then, the accessing cost will depend on the total usage of the link at that time and the load at the link, $ e_{mn} $ is given by $\psi_{e_{mn}}(t_i)= \varsigma_{e_{mn}}(\sum_{r : e_{mn} \in E} x_r(t_i))$ where, $\varsigma_{e_{mn}}(\bullet)$ is a costing function and is growing if it comes in large number of user's path. $\varsigma_e(y)$ is given by 
$ \varsigma_{e_{mn}}(y)= c_{e_{mn}}.(y/C_{e_{mn}})^\omega $ where, $c_{e_{mn}}$ is normalizing constant. Now, consider the following system of differential equation for getting optimal data rate ($ x^* $)\\
\begin{equation}
\frac{dx_r(t_i)}{dt_i}= \vartheta_r(\mathcal{P}_r(t_i)-x_r(t_i)\sum_{e_{mn} \in r}\psi_{e_{mn}}(t_i)) \label{e4}
\end{equation}
Here, $ \vartheta_r $ is proportionality constant. Each user first computes it's willingness to pay as $\mathcal{P}_r(t_i)$ then, it adjusts its rate based on the response provided by the links in the network and trying to balance its willing to pay and total price. Eq. \eqref{e4} consists of two components: a steady increase in the rate proportional to $\mathcal{P}_r(t_i)$ and steady decrease in the rate proportional to the response $ \psi_{e_{mn}}(t_i) $ is provided by the network.

\section{Time Varying Disassortative Communication network model }
For the smaller value of the packet generation rate, $ \lambda $ system remains in free flow state as every packet is getting delivered. But with the increasing value of $ \lambda $, a point is reached where system converts into congested phase and this point of phase transition is known as critical packet generation rate, $ \lambda_c $. The value of $ \lambda_c $ is affected by the topology of the network. Therefore, in this paper, disassortative TVCN (DTVCN) model is proposed to achieve maximum value of $ \lambda_c $. As we know that the positive correlation increases congestion in the communication network. Hence, in the proposed model, if a new node will appear at any time instant $ t_{i+1} $then it will be attached to the network which exists at $ t_i $. A new node may be attached to the existing nodes by preferring higher degree and disassortativity with the neighbors. As DDC is directly proportional to the congestion of a node hence, some fractions of the congested links are rewired. While some fraction of anti-preferential and correlated links are removed from the network. In this way, congestion is minimized and we get the higher value of critical packet generation rate, $ \lambda_c $.

The scale free (BA)model assumes that the new node will prefer to attach with the nodes in the existing network based on the value of the degree of the existing node. But, in most of the networks, this assumption may not be true, as nodes can't acquire links unconditionally. There is a limit on packet forwarding rate i.e., the capacity of the node. If more links are attached to a node then, it may happen that the node will be a part of a large number of user's shortest paths and leads to congestion in the network. Therefore, we modified the probability to account for the congestion and introduced correlation as a multiplicative factor to the preferential attachment probability. Probability $ \Pi $ that a node  $ i $ will be selected through preferential attachment is proportional to its degree and is given by,$ \frac{k_i}{\sum_{j \in N} k_j} $ whereas the probability $ \Pi' $ of selecting node $ i $ with anti-preferential attachment is given by, $ \frac{1}{|N|-1} \left(1-\frac{k_i}{\sum_{j \in N} k_j}\right) $. Like BB model \cite{BB} an the model \cite{tee2016}, DTVCN model provides conditioning on the probability of preferential attachment with the probability of node's disassortativity with its neighbors and is defined as:
\begin{equation}
\Pi_i^r =  \frac{k_i}{\sum_j k_j} (1 + \zeta_i)\label{e5}
\end{equation}
Where, $ \zeta_i = \frac{r_{deg}(i)}{\sum_{n=1}^{|Ne(i)|} r_{deg}(i,n)} $, $ r_{deg}(i) = min \mbox{ } r_{deg}(i,n), \forall {n: n \in Ne(i)} $, $ Ne(i)= $ Neighbors of node $ i $ and $ 0 \leq r_{deg}(i,n)+1 \leq 2$
The probability $ \Pi_i^{r'} $ of selecting a node $ i $ with anti-preferential attachment and high correlation with its neighbors is given by, $ \left( 1-\frac{k_i}{\sum_j k_j}\right) (1-\zeta_i) $.


Algorithmic steps are given for addition, rewire and removal of links in the proposed DTVCN model is given in  (Algorithm \ref{al1}).\\
\begin{algorithm}[htb!]
\begin{algorithmic}[1]
\STATE \textbf{Input:}  A number of nodes $ (n_0) $ for creating seed network and $ T $.
\STATE \textbf{Output:} Time varying data communication networks.  
\WHILE{$ t  \leq T $}
\STATE Expand the network with one node at each time instant $ t $ and select $ M (\leq t)$, $ \beta $ and $ \gamma $ and calculate $ f_{add} (t), f_{rewire}(t) $ and $f_{delete}(t)$.   
\FOR{$ x $ : 1 to $ f_{add}(t) $}
\STATE  Choose a node in existing network with probability, $ \Pi_i^{r} $.
\ENDFOR
\FOR{ $ y $ : 1 to $f_{rewire}(t)$}
\STATE Remove a link of a node with probability, $ \Pi_i^{r'} $ and attach to the node having higher probability value, $ \Pi_i^{r} $.
\ENDFOR
\FOR{$ z $ : 1 to $f_{delete}(t)$}
\STATE Remove the infrequently used and correlated link of a randomly selected node with probability, $ \Pi_i^{r'} $.
\ENDFOR
\ENDWHILE\label{endtimer}
\end{algorithmic}
\caption{Time varying disassortative communication network model}
\label{al1}
\end{algorithm}
\vspace*{-0.5 cm}

After studying the scaling properties of the time varying disassortative network model, the SF behavior is found. A mean field theoretical approach \cite{bara} is used to anticipate the growth dynamics of distinct nodes, which is later applicable for analytical computation of connectivity distribution and scaling exponents. The network is growing with time and a node will acquire a link if its degree is high as well disassortative with its neighbors. Degree $ k_i $ of the node $ i $ is changing continuously with time, so probability $ \Pi(k_i)  $ is interpreted as rate of change of $ k_i $. At each time stamp, fractions $ \beta $ and $ \gamma $ decide the number of links chosen for expansion (addition), rearrangement and removal, and are deciding parameters of degree for the node $ i $. The analysis may be given as:
\begin{enumerate}
\item A fraction $ \beta $ of the number $ M_t $ links are newly added links at time $ t $.
\begin{equation}
\left( \frac{dk_i}{dt}\right)_{add} = \beta M_t \frac{k_i}{\sum_j k_j} (1 + \zeta_i)\label{e6}
\end{equation}
Here, $ \zeta_i = \frac{r_{deg}(i)}{\sum_{n=1}^{|Ne(i)|} r_{deg}(i,n)} $, $ r_{deg}(i) = min \mbox{ } r_{deg}(i,n), \forall {n: n \in Ne(i)} $, $ Ne(i)= $ Neighbors of node $ i $ and $ 0 \leq r_{deg}(i,n)+1 \leq 2$

The effect of addition of newly added link on the rate of change of degree of a node, $ i $ with time is written on the left hand side of the equation and on the right side $ \beta M_t $ links are formed by using preferential attachment and the value of $ \zeta_i $.  The rate of change of degree of a node, $ i $ is directly proportional to its own degree as well as on the correlation with its neighbor.  
\item A fraction $ \gamma (1-\beta) M_t $ links are re-arranged at time $ t $. 
\begin{equation}
\begin{aligned}
\left( \frac{dk_i}{dt}\right)_{R} ={}& \gamma (1-\beta) M_t[ \frac{1}{n_0 + t} + \left( 1- \frac{1}{n_0 + t} \right) \frac{k_i}{\sum_j k_j} (1+\zeta_i)\\ 
& - \frac{1}{n_0 + t} \left( 1-\frac{k_i}{\sum_j k_j}\right) (1-\zeta_i) ] \label{e7}
\end{aligned}
\end{equation}
In the time varying networks few links are getting rewired and the degree of node $ i $ depends on three terms: first term shows random selection of nodes, second term corresponds to linking with other existing nodes having high preferential attachment probability and higher anti-correlation with neighbor nodes and third term shows removal of link having low preferential attachment value and higher correlation with its neighbors. 
\item A fraction $ (1-\gamma) (1-\beta) M_t $ links are removed from the network at time $ t $.
\begin{equation}
\begin{aligned}
\left( \frac{dk_i}{dt}\right)_{delete} ={}& -(1-\gamma)(1-\beta)M_t [\frac{1}{n_0+t} + \left(1-\frac{1}{n_0 + t} \right) \left( 1-\frac{k_i}{\sum_j k_j}\right) \\
& (1- \zeta_i) \frac{1}{n_0+t} ] \label{e8}
\end{aligned}
\end{equation}
Removal of links affects the degree of node $ i $ and it is shown in above equation. First term shows random selection of a node and that link will be removed which has low preferential attachment value as well as highly correlated with its neighbors.
\end{enumerate}

At time $ t $, the sum of degrees of nodes in the network will be:
\begin{equation*}
\sum_j k_j = 2t \left[\beta M_t + (1-\beta)\left(\gamma M_t - (1-\gamma) M_t \right)\right]
		= 2M_tt\left[ \beta + (1-\beta)(2\gamma-1)\right]  
\end{equation*} 
Let $ c = \beta + (1-\beta)(2\gamma-1) $.\\
Now, combining the Eqns. \eqref{e6} to \eqref{e8} we get the change in degree of node $ i $ with respect to time $ t $.

\begin{equation}
\begin{aligned}
\frac{dk_i}{dt} ={} & \frac{\beta k_i}{2ct}(1 + \zeta_i) + \\
& \gamma(1-\beta)M_t \left[ \frac{1}{t} + \frac{k_i (1 + \zeta_i)}{2cM_tt} - \frac{k_i(1 + \zeta_i)}{2cM_tt^2} - \frac{(1- \zeta_i)}{t} + \frac{k_i (1 - \zeta_i)}{2cM_t t^2} \right] \\
& - (1-\beta)(1-\gamma)M_t \left[ \frac{1}{t} + \frac{(1- \zeta_i)}{t} - \frac{(1- \zeta_i)}{t^2} - \frac{k_i (1-\zeta_i)}{2cM_tt^2} + \frac{k_i(1-\zeta_i)}{2cM_tt^3} \right]\\
={} & \left[\frac{(1+ \zeta_i)}{2c} \left(\beta + \gamma(1-\beta)\right)\right]\frac{k_i}{t} + 
 \left[\left((2\gamma- 2 + \zeta_i)(1-\beta) M_t\right) \right]\frac{1}{t}\\
&\mbox { (for large  t ) } \label{e9}
\end{aligned}
\end{equation}
Let, $ \mathcal{K}_1 = \frac{(1+ \zeta_i)}{2c} \left(\beta + \gamma(1-\beta)\right)  $ and 
$ \mathcal{K}_2 = (2\gamma- 2 + \zeta_i)(1-\beta) M_t  $\\
Hence, the equation Eq. \eqref{e9} can be re-written as\\
\begin{equation}
\frac{dk_i}{dt} = \mathcal{K}_1 \frac{k_i}{t} + \mathcal{K}_2 \frac{1}{t} \label{e10}
\end{equation}
The solution of the Eq. \eqref{e10} is derived by considering the initial conditions that node $ i $ appears at time $ t_i $ with $ M_t $ connections $ k_i(t_i) = M_t $ and is given as:
\begin{equation*}
\begin{aligned}
k_i(t)={} & -\frac{\mathcal{K}_2}{\mathcal{K}_1} + \left( k_i(t_i) + \frac{\mathcal{K}_2}{\mathcal{K}_1} \right) \left( \frac{t}{t_i} \right) ^ {\mathcal{K}_1} \\
   ={}   & -\frac{\mathcal{K}_2}{\mathcal{K}_1} + \left( M_t + \frac{\mathcal{K}_2}{\mathcal{K}_1} \right) \left(\frac{t}{t_i} \right) ^ {\mathcal{K}_1}
\end{aligned}
\end{equation*}
At each time stamp, a new node is added into the existing network hence, network probability density for $ t_i $ at previous time
\begin{equation*}
P_i(t_i) = \frac{1}{n_0 + t}
\end{equation*}
The probability of a node that has total $ k_i(t) $ connections and which is smaller than $ k $, $ P(k_i(t) < k) $, can be written as,
\begin{equation*}
\begin{aligned}
P(k_i(t) < k) = {} & P(-\frac{\mathcal{K}_2}{\mathcal{K}_1} + \left( M_t + \frac{\mathcal{K}_2}{\mathcal{K}_1} \right) \left(\frac{t}{t_i} \right) ^ {\mathcal{K}_1} < k)\\
={}  & P \left( t_i > \left( \frac{M_t + \frac{\mathcal{K}_2}{\mathcal{K}_1}}{k+\frac{\mathcal{K}_2}{\mathcal{K}_1}} \right) ^ {\frac{1}{\mathcal{K}_1}} t \right)\\
={} & 1 - P \left( t_i \leq \left( \frac{M_t + \frac{\mathcal{K}_2}{\mathcal{K}_1}}{k+\frac{\mathcal{K}_2}{\mathcal{K}_1}} \right) ^ {\frac{1}{\mathcal{K}_1}} t \right)
\end{aligned}
\end{equation*}
The probability density function of $ k $ is $ P(k) $ and can be written as,
\begin{equation}
\begin{aligned}
P(k) = {} & \frac{\partial P(k_i(t) < k) }{ \partial k}\\
={} & \frac{t}{\mathcal{K}_1(n_0+t)} \left( M_t + \frac{\mathcal{K}_2}{\mathcal{K}_1} \right) ^ {\frac{1}{\mathcal{K}_1}}\left( k + \frac{\mathcal{K}_2}{\mathcal{K}_1} \right) ^ {-(1+\frac{1}{\mathcal{K}_1})} \label{e11}
\end{aligned}
\end{equation}
Now, assume $ \alpha = -(1+\frac{1}{\mathcal{K}_1}) $. \\
The value of the scaling exponent of the real world network lies between $ 2 $ and $ 3 $, hence, exponent, $ \alpha $ of the proposed dynamic communication network model must lie within $ 2 < \alpha \leq 3  $. Constraint on $ \alpha $ will be fulfilled if $ 0.5 < \mathcal{K}_1 < 1 $. The value of $ \mathcal{K}_1 $ is dependent on the parameters, $ \beta $, $ \gamma $ and $ \zeta_i $. $ \zeta_i $ is dependent on the value of $ r_{deg}(i) $ and is scaled from $ -1 < r_{deg}(i) \leq 1 $ to $ 0 < r_{deg}(i) \leq 2 $. The complete network is connected therefore, node $ i $ must be connected with at least one other node and hence, maximum value of $ \zeta_i = 1 $ and the range of $ \zeta $ is $ 0 < \zeta_i \leq 1$. From Table $ \ref{tab1} $, for different values of parameters, $ \beta $, $ \gamma $ and $ \zeta $, the power law exponent, $ \alpha $ lies between $ 2 $ and $ 3 $ as shown in Table \ref{tab1}. As $ P(k) $ is always positive so, $ M_t + \frac{\mathcal{K}_2}{\mathcal{K}_1} $ should be positive.

\begin{table*}[!htb]
\caption{The value of power law exponent, $ \alpha $ obtained for different values of $ \beta $, $ \gamma $ and $ \zeta $.} 
 \center
 \adjustbox{max height=\dimexpr\textheight-5.5cm\relax,
           max width=\textwidth}{
\begin{tabular}[1]{l| c| c| c}
\hline\hline
$ \zeta $ & \textbf{$ \beta $} & \textbf{$ \gamma $} & \textbf{$ \alpha $} \\ [0.5ex]
\hline\hline
$ 0.0100 $ & $ 0.5100 $ & $ 0.5100 $ & $ 2.3545 $ \\[0.25ex]
\hline
$ 0.3333 $ & $ 0.6000 $ & $ 0.6000 $ & $ 2.2143 $ \\[0.25ex]
\hline
$ 0.5000 $ & $ 0.7500 $ & $ 0.7500 $ & $ 2.2444 $ \\[0.25ex]
\hline
$ 1.0000 $ & $ 0.9999 $ & $ 0.9999 $ & $ 3.0000 $ \\[0.25ex]
\hline
\end{tabular}
\label{tab1}
}
\end{table*}

\section{Time Varying Routing Strategy}

The degree is a fundamental quantity to measure topology and influence of the node in the SF network. While betweenness centrality (BC) is used to measure the influence of a node in communication between each pair of nodes in the network. Betweenness centrality (BC) is another important measure which is used to find the extent to which a node lies on the shortest paths between another pair of nodes. BC of a node $ i $ is, $ g_i = \sum_{s \neq i \neq d} \frac{n_{s,d}^{i}}{n_{s,d}} $ where, $ n_{s,d}^{i} $ is the number of the shortest paths from node $ s $ to $ d $ passing through node $ i $ and $ n_{s,d} $ is the total number of shortest paths from node $ s $ to node $ d $. A node with high BC may have considered more influential within the network. A large number of users are sending data through the node hence, removal of the node will disrupt the communication in the network. The BC distribution follow power law and BC is related to degree as $ g \sim k^{(\frac{\alpha-1}{\delta-1})}$ where, $ \delta $ is BC exponent. This relation shows that node with a larger degree will also be influential in communication also. Hence, we may say that the behavior of BC-BC correlation would be similar to DDC.  

The structure of the network affects the value of critical packet generation rate, $ \lambda_c $. For a small value of $ \lambda_i $ of a node $ i $, the number of packets in the network is small hence, all the data will be processed. If all the packets are sent through the shortest paths in the network to their destinations then, some nodes may appear frequently in the formation of shortest paths. The data forwarding capacities of these nodes got reduced and it increases congestion locally in all the paths where it appeared and gradually increase overall congestion. Multiple users want to establish a connection between a source node, ($s$) and a destination node, ($d$) in large communication networks. Multiple shortest paths may exist, $ \sigma_z(s \rightarrow d) $ from $ s $ to $ d $, where, $ z = (1,2, ..., \chi) $ and $ \chi $ is the total number of shortest paths between $ s $ and $ d $. A node may highly correlate with its neighbors in the shortest path. Data sending through the shortest paths may lead to the congestion in the network. Therefore, it is important to investigate the shortest path $ \sigma(s \rightarrow d) $ between the user pairs such that the overall BC-BC correlation of the nodes that appear in the path should be minimum.
\begin{equation*}
W_g[\sigma_z(s \rightarrow d)] = \sum_{v: v \in \sigma_z(s \rightarrow d)} r_g(v)
\end{equation*} 
Where, $ r_g(v) = min \mbox{ } r_g(v,n), \forall {n: n \in Ne(v)} $, $ Ne(v)= $ Neighbors of node $ v $ and $ 0 \leq r_g(v,n)+1 \leq 2$
Hence, we want to find a path whose nodes should not be the part of hub nodes. Hence, it is defined by,
\begin{equation*}
min \mbox{  } \{\forall z : W_g[\sigma_z(s \rightarrow d)]\}
\end{equation*}

Each user provides the information about the source node, $ s $ and destination node $ d $ and accordingly $ (s,d) $ pairs are generated. Increments in the number of users lead to the network into a congested state. Therefore, our aim is to find efficient routing paths such that the maximum number of users are getting benefited with a unique stable value of the optimal data sending rate $ x_r^* $ and corresponding convergence vectors will be $ x^*={x_r^*, r \in R}$ using equation Eq. (\ref{e4}). Shortest path $ (s \rightarrow d) $ with lowest $ W_g $ is assumed as efficient. After finding such paths of all the users, optimal data sending rate $ x^* $ of each user is calculated by using rate allocation equation Eq. (\ref{e4}). A detailed description of selecting the shortest path with lowest (highest) $ W_g $ along with optimal rate is given in Algorithm \ref{al2}. 

\begin{algorithm}[H]
\begin{algorithmic}[1]
\STATE \textbf{Input:} All source destination $ (s,d) $ pairs $ \mathcal{N}_{sd} $ in the network, designed through Steps in Section 3.1, $ a $ and $ b $ such that $ a > 0 $ \& $ 0 < b < 1 $.
\FOR {$ i:=  1 \mbox{ to } length(\mathcal{N}_{sd})$}
\STATE Evaluate all shortest paths $ \chi_i $ of user $ i $.
\IF{ $  \chi_i  > 1 $}
\STATE Select shortest paths $ \chi_i(s \rightarrow d) $  having maximum and minimum values of $ W_g [\sigma_m(s \rightarrow d)] $.
\ELSE
\STATE Calculate $ W_g[\sigma(s \rightarrow d)] $.   
\ENDIF
\ENDFOR
\FOR { $ i := 1 \mbox{ to } length(\mathcal{N}_{sd}) $ }
	 \FOR { $ d := 1 \mbox{ to } length(\mathcal{N}_{sd}(i))$}
        \STATE Update network feedback $ \psi_d $ for each element $ d $.
     \ENDFOR
        \STATE $ x_r = min(x_{\mathcal{N}_{sd}(i)}) $;
        \STATE $ A(r) = rand(1,10) $;
        \STATE $ \mathcal{P}_r = x_r * (\frac{a}{x_r+ b}) $;
        \STATE $ \psi_r = \psi_d \{ \forall d : d \in \mathcal{N}_{sd}(i)\}$;
\ENDFOR
\STATE Use the value of $ x_r $, $ A(r) $, $ \mathcal{P}_r $ and $ \psi_r $ to find the rate of convergence of each user.  
\end{algorithmic}
\caption{Finding shortest path having lowest (and highest) $ W_g $ and $ x^* $ for each user }
\label{al2}
\end{algorithm}

\section{Simulation and Results}
For dynamics of DTVCN model, the simulation starts by establishing the infrastructure of the network followed by algorithmic steps in Algorithm \ref{al1}. In this paper, the parameters are set with the value as seed node $ n_0 = 5 $, number $ M (\leq t) $, fraction of newly added links $ \beta $ range in $ (0,1) $, fraction of rewired links $ \gamma $ is in the range of $ (0.5, 1) $, with network size ranging from $ |N| = 10^3 $ to $ |N| = 10^4 $. Any node can be included in the user's $ (s,d) $ sets or may participate in routing also. For simplicity, we have assumed the capacity of all nodes is equal. At each time stamp, degree of the nodes will be different, hence, flow of data through the nodes as well as links change accordingly. The range of degree degree correlation coefficients, $ r_{deg} $ and BC-BC coefficient, $ r_g $ are scaled from $ (-1,1) $ to $ (0,2) $. Utility function, $ U_r(x_r) $ is a concave function, $U_r(x_r) = a \mbox{  }log(x_i + b) $ 

\begin{figure*}[!htb]
\begin{center}
$\begin{array}{ccc}
\includegraphics[width=0.33\linewidth, height=1.75 in]{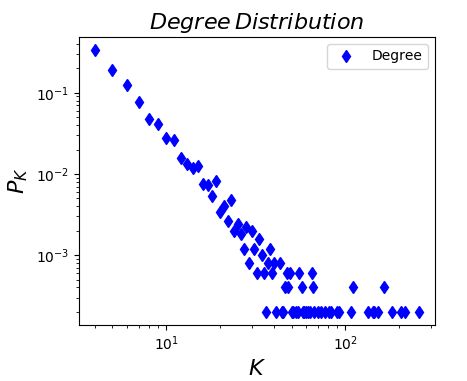}&
\includegraphics[width=0.33\linewidth, height=1.75 in]{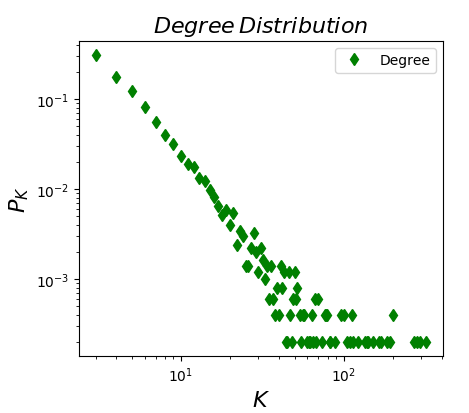} &
\includegraphics[width=0.33\linewidth, height=1.75 in]{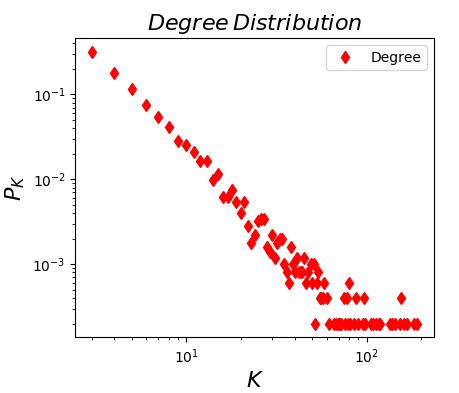}
\\ 
\mbox{(a)} & \mbox{(b)} & \mbox{(c)}  \\
\end{array}$
\caption{Degree distribution of the network when $ |N| = 5 \times 10^3 $, correlation coefficient, $ r_{deg} $ and Power law exponent, $ \alpha $ for all the models, BA model,TVCN model and DTVCN model are \textbf{(a)} $ r_{deg}= -0.03708 \mbox{ , } \alpha = 2.6812 $ , \textbf{(b)} $ r_{deg} = -0.02648 \mbox{ , } \alpha = 2.2520$  and \textbf{(c)} $ r_{deg} = -0.02239 \mbox{ , }\alpha = 2.2432 $ respectively.}
\label{f1}
\end{center} 
\end{figure*}

Degree distribution of all the three models i.e., BA model, TVCN model, and the proposed DTVCN model is shown in Fig. \ref{f1}. The network generated through all the models follow power law degree distribution and the exponent, $ \alpha $ is in the range of $ (2,3] $ hence, scale free in nature. The values of degree degree correlation coefficient, $ r_{deg} $ shows that the generated communication networks are disassortative. The network designed through the BA model is more disassortative than the other two models.  
 
In Table \ref{tab1}, various network properties such as clustering coefficient, diameter, average path length, critical packet generation rate, $ \lambda_c $, power law exponent, $ \alpha $, rich club coefficient (RC) are studied. The reason behind selecting all these properties as opposed to other possible properties is because they are properties of the network as a whole and not just of the individual node. This allows for analysis of how the whole network changes and not just the structure around some particular node. The size of the network, $ |N| $ is varying from $ 2 \times 10^3 $ to $ 1 \times 10^4 $. All these network properties are studied for the network generated by different models, Barabasi-Albert (BA)\cite{bara}, TVCN \cite{kumari2017}, and the proposed DTVCN model. BA model considers the only addition of links with preferential attachment, TVCN model considers the addition and rewiring of links with preferential attachment while removal of the link is based on anti-preferential attachment. The proposed DTVCN model takes care of congestion in the network by introducing DDC of a node along with the preferential attachment. We know that there is a critical packet generation rate $ \lambda_c $, below which the network traffic is free but above which traffic congestion occurs. By calculating the value of $ \lambda_c $ for all three models, the proposed DTVCN model is found to be more tolerant to congestion. In the proposed DTVCN model, the value of the highest BC is lesser than the other models hence, it gives approximately $ 18.6075\% $ and $ 25.6950\% $ higher value of $ \lambda_c $ than TVCN model and BA model respectively. This shows that each node can generate and forward more packets and makes the network more congestion tolerant. The clustering coefficient of TVCN model is the highest while, BA model has the least value. Clustering coefficient decreases with increasing value of network size, $ |N| $. The length of average Shortest path of BA model is greater than others while TVCN model gives the lowest value. The value of power law exponent, $ \alpha $ is in the range of $ 2 < \alpha \leq 3 $. Rich club phenomenon is characterized when the hubs are on average more intensely interconnected than the nodes with smaller degrees. When the nodes with degree larger than $ k $ tend to more connected than the nodes with degree smaller than $ k $. Presence of rich club indicates robustness against hub failure but increases load and congestion on the connecting link between two hubs. Most of the users want to send data through shortest paths i.e., through hub nodes and congestion at hub nodes will reduce the data forwarding capacity and efficiency of the networks. The proposed DTVCN model takes care of congestion and gives the lowest value of RC.    

\begin{table*}[!htb]
\caption{Network size $ |N| $, clustering coefficient, diameter, average path length, power law exponent, $ \alpha $, critical packet generation rate, $ \lambda_c $ and rich club coefficient, $ RC $ for all the models; proposed DTVCN, TVCN and BA when, the network size, $ |N| $ varying from $  = 2 \times 10^3 $ to $ 1 \times 10^4 $.} 
 \center
 \adjustbox{max height=\dimexpr\textheight-5.5cm\relax,
           max width=\textwidth}{
\begin{tabular}[1]{l| c| c| c| c| c| c| c}
\hline\hline
 & Network & clustering &  Diameter & Average &  & $ \lambda_c $ & $ RC $  \\ [0.5ex]
$ |N| $ & Model & coefficient  & & Path Length & ($ \alpha $) & &  \\
\hline\hline
 & DTVCN & $ 0.0324 $ & $ 6.0000 $ & $ 3.4220 $ & $ 2.2129 $ & $ 0.6454 $ & $ 0.0605 $ \\[0.25ex]
$ 2000 $ &  TVCN & $ 0.0456 $ & $ 6.0000 $ & $ 3.3055 $ & $ 2.2452 $ & $ 0.5014 $ & $ 0.0993 $ \\[0.25ex]
& BA & $ 0.02515 $ & $ 5.0000 $ & $ 3.3635 $ & $ 2.6778 $ & $ 0.3835 $ & $ 0.1397$  \\[0.25ex]
\hline
 &  DTVCN & $ 0.0180 $ & $ 6.0000 $ & $ 3.6525 $ & $ 2.2432 $ & $ 0.4140 $ & $ 0.0383 $ \\[0.25ex]
$ 5000 $ & TVCN & $ 0.0286 $ & $ 6.0000 $ & $ 3.5256 $ & $ 2.2520 $ & $ 0.2472 $ & $ 0.0696 $ \\[0.25ex]
 & BA & $ 0.0123 $ & $ 6.0000 $ & $ 3.6685 $ & $ 2.6812 $ & $ 0.1950 $ & $ 0.0582 $ \\[0.25ex]
\hline
 &  DTVCN & $ 0.0141 $ & $ 6.0000 $ & $ 3.6525 $ & $ 2.2498 $ & $ 0.3930 $ & $ 0.0304 $ \\[0.25ex]
$ 7000 $ & TVCN & $ 0.0226 $ & $ 6.0000 $ & $ 3.6068 $ & $ 2.2534 $ & $ 0.1880 $ & $ 0.0635 $ \\[0.25ex]
 & BA & $ 0.0094 $ & $ 6.0000 $ & $ 3.7701 $ & $ 2.6843 $ & $ 0.1437 $ & $ 0.0522 $ \\[0.25ex]
\hline
 &  DTVCN & $ 0.0113 $ & $ 7.0000 $ & $ 3.8249 $ & $ 2.2533 $ & $ 0.3675 $  & $ 0.0274 $\\[0.25ex]
$ 10000 $ & TVCN & $ 0.0181 $ & $ 6.0000 $ & $ 3.6961 $ & $ 2.2544 $ & $ 0.1390 $ & $ 0.0548 $ \\[0.25ex]
 & BA & $ 0.0056 $ & $ 7.0000 $ & $ 4.3006 $ & $ 2.5723 $ & $ 0.1046 $ & $ 0.0446 $\\[0.25ex]
\hline
\end{tabular}
\label{tab1}
}
\end{table*}

\begin{figure}
\minipage{0.45\textwidth}
    \includegraphics[width= \linewidth, height=2 in]{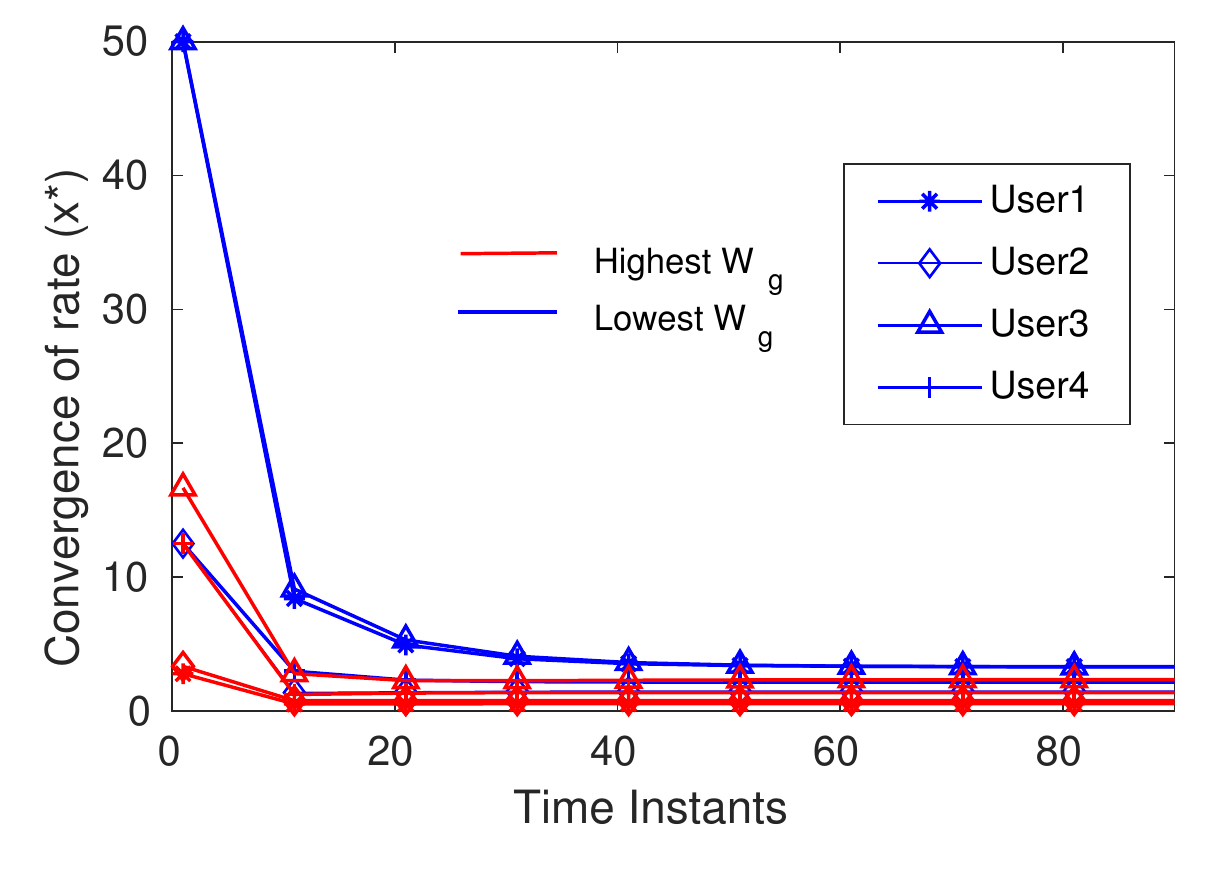}
    \center{(a.)}
    \label{fig:my_label}
\endminipage\hfill
\minipage{0.45\textwidth}
    \includegraphics[width= \linewidth, height=2 in]{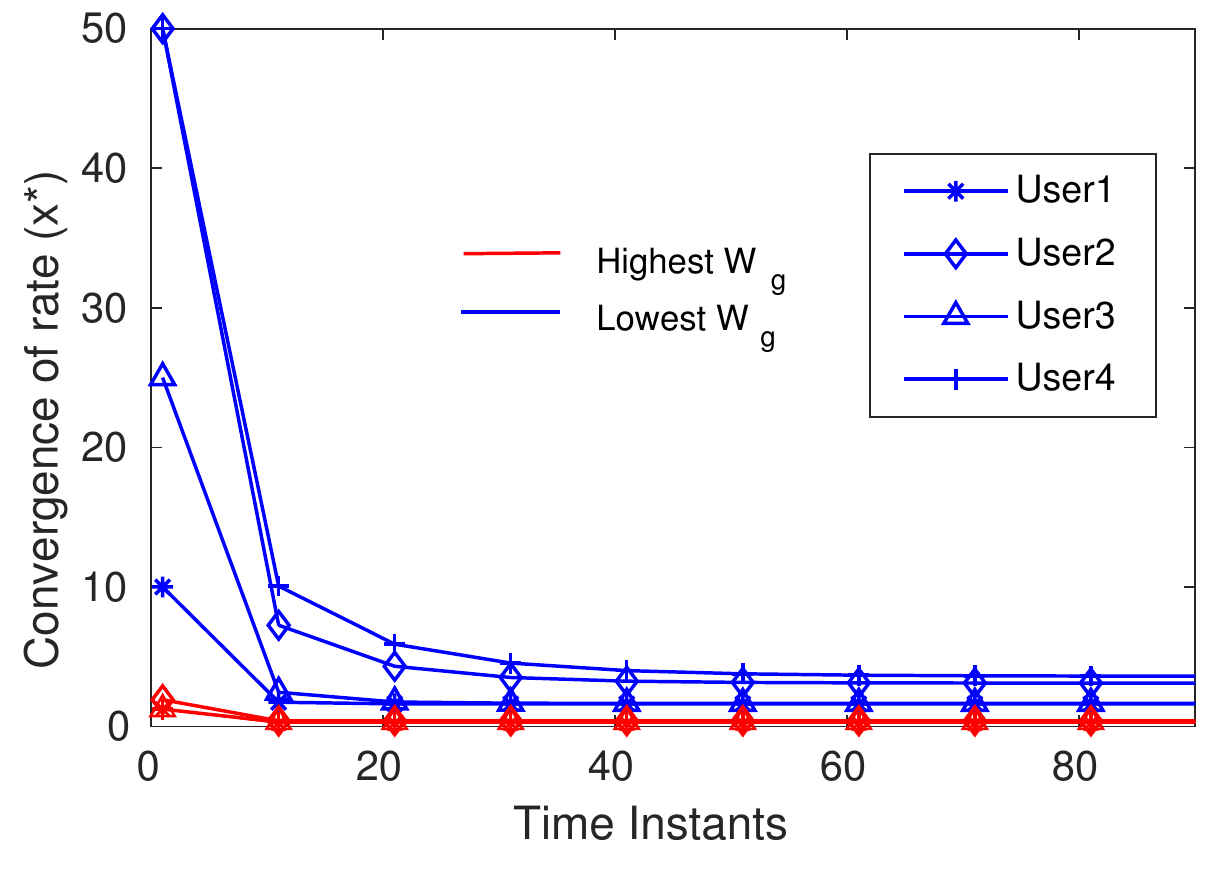}
    \center{(b.)}
    \label{fig:my_label}
\endminipage\hfill
\minipage{0.45\textwidth}
	\begin{center}
    \includegraphics[width = \linewidth, height=2 in]{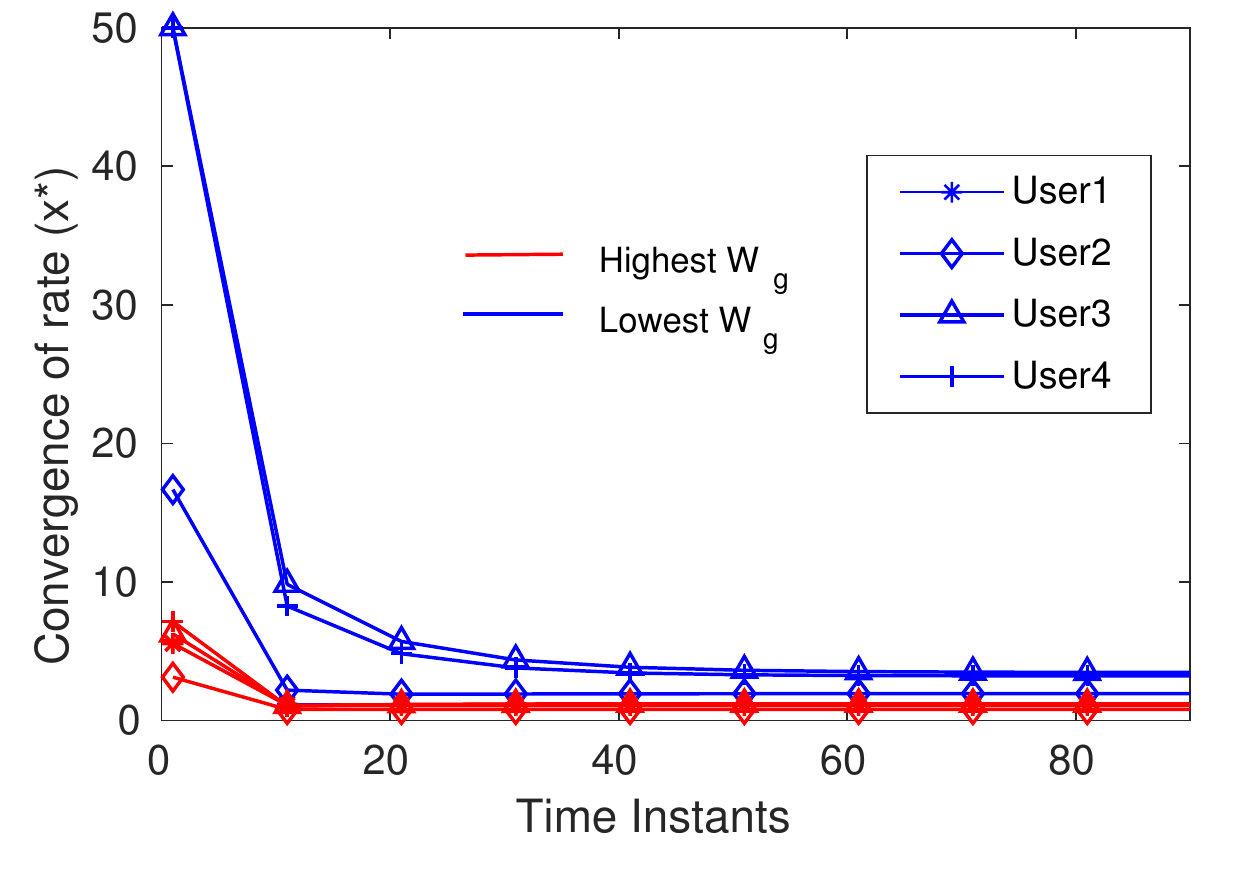}   
    {(c.)}
    \end{center}
    \endminipage
\caption{Convergence of user's data rate, $ x^* $, when the size of the network, $ |N| = 2 \times 10^3 $ for all the models, \textbf{(a)} TVCN model, \textbf{(b)} BA model and \textbf{(c)} DTVCN model respectively.}
\label{f2}

\end{figure}

Each user may send its data through one of the shortest paths. In this paper, two shortest paths are considered on the basis of the value of $ W_g $. As network resources are shared among a large number of users hence, each user, $ r $ has to compromise on data rate, $ x_r $. User's rate depends on the demand of resources appearing
in the shortest route. If demand is high, the data rate will be less. User $ r $ first computes its willingness for pay, $\mathcal{P}_r(t_i)$ then, it adjusts its rate based on the congestion and the response provided by the links appeared in the route. An optimal data sending rate, $ x_r^*$ of each user, $ r $ is obtained by using rate control theorem given in Eq. \eqref{e4}. Fig. \ref{f2} shows the convergence of the data rate of four users for the network proposed through all the three models. Optimal data rate, $ x^* $ is minimum (or maximum) for maximum (or minimum) value of $ W_g $.

\begin{figure*}[!htb]
\begin{center}
$\begin{array}{cc}
\includegraphics[width=0.5\linewidth, height=2 in]{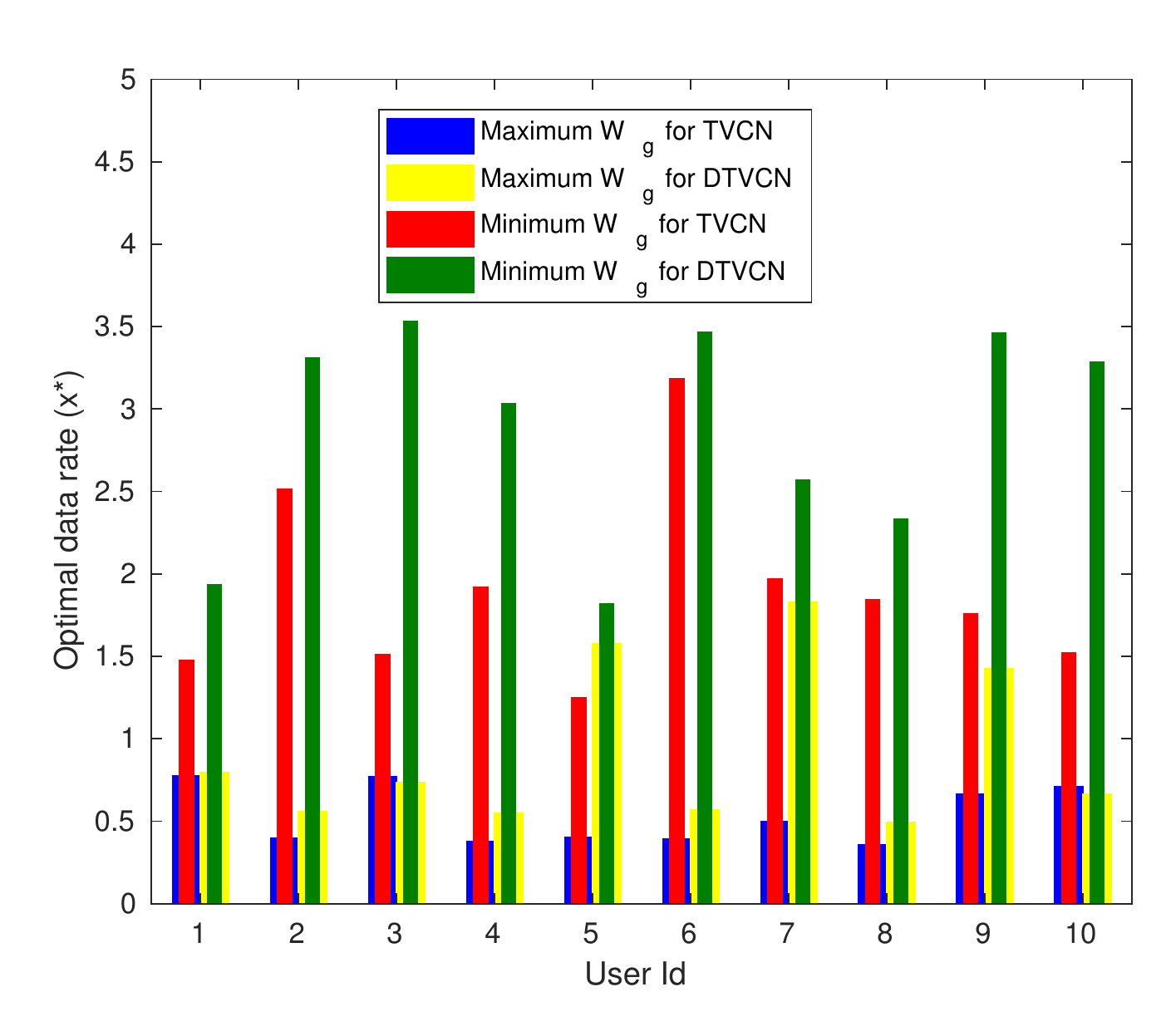} &
\includegraphics[width=0.5\linewidth, height=2 in]{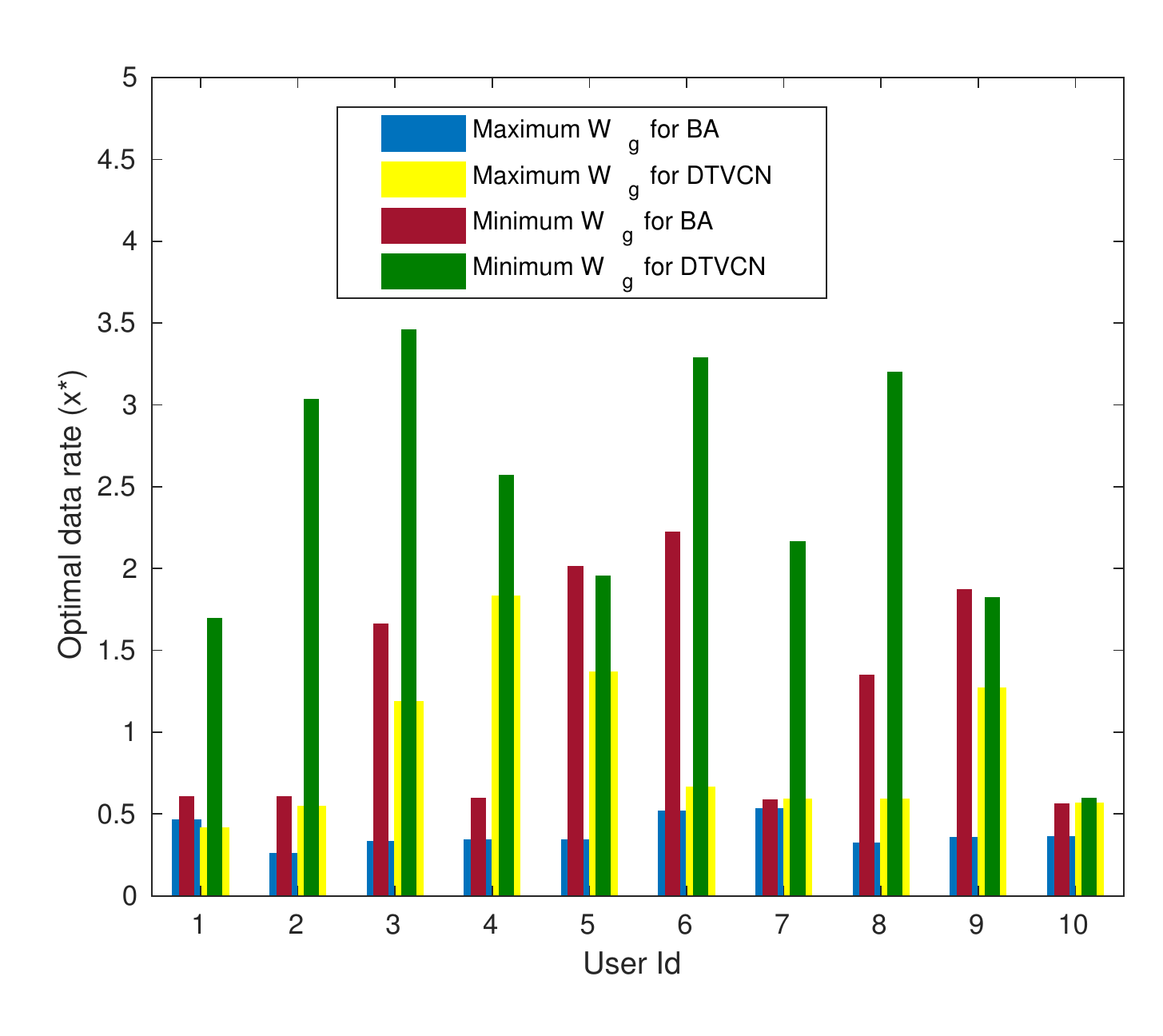}\\ 
\mbox{(a)} & \mbox{(b)}  \\
\end{array}$
\caption{User's optimal data rates $ x^* $ at time $ t = 2 \times 10^3 $ by considering shortest path with maximum \& minimum value of betweenness correlation( $  W_g[n_{s,d}] $) for \textbf{(a.)} TVCN vs. DTVCN and \textbf{(b.)} BA vs. DTVCN.}
\label{f3}
\end{center} 
\end{figure*}

Size of the network, $ |N| $ is $ 2 \times 10^3 $ and total $ 100 $ users want to access the networks. Optimal data sending rate of all the users with highest and lowest $ W_g $ are calculated for BA model, TVCN model, and DTVCN  model. The result of optimal data rates ($ x^* $) of $ 10 $ users is shown in Fig.\ref{f3}(a) and Fig. \ref{f3}(b). The path of the user will depend on the available options and may be different. Therefore, data forwarding rate may also vary accordingly. The lowest value of $ W_g $ shows least correlation of the nodes appearing in the user's route with its neighbors. As congestion increases with correlation hence, the data rate of the user having lowest or highest value for the $ W_g $ is always maximum or minimum. User's route depends on the structure of the network. The size of the network, $ |N| $ and link density for all the models; BA, TVCN, and DTVCN are same but topologies are different. Most of the users get a higher value of $ x^* $ for both maximum (or minimum) value of $ W_g $ in the network designed by the proposed modeling technique i.e., DTVCN as shown in Fig. \ref{f3}(a.) and Fig. \ref{f3}(b.). Fig. \ref{f3}(a.) shows that the average of the users' data rate, $ x^* $ is $ 2.0981 $ and $ 0.7117 $ for lowest and highest value of $ W_g $ respectively in TVCN model. While, by using DTVCN model, the value of average users' data rate, $ x^* $ is $ 2.4258 $ and $ 1.0322 $ for lowest and highest value of $ W_g $ respectively. In Fig. \ref{f3}(b.), the value of average user' data rate, $ x^* $ is $ 2.0024 $ and $ 0.5103 $ for lowest and highest value of $ W_g $ respectively for the network proposed by using BA model. 

\section{Conclusions and Future Work}
In the present work, the effect of DDC to the congestion against the dynamic process on SF network is studied. As a result, it is found that consideration of disassortativity in network modeling performs better for data communication as compared to the other models i.e., BA model and TVCN model. The value of critical packet generation rate, $ \lambda_c $ could explain why the network is being modeled by considering DDC along with the preferential attachment. The value of $ \lambda_c $ is inversely proportional to the highest BC of the network and BC of a node is proportional to its degree.  We also showed that the value of power law exponent, $ \alpha $ decreases as compared to other models. Hence, the degree of the maximum degree node in the network will scale down than the BA model and TVCN model. The value of rich club coefficient shows that the hubs are on average less intensely interconnected than the hubs in other two models. Shortest path with lowest or highest $ W_g $ are chosen for routing and observed that the value of $ x_r* $ will always be the maximum or minimum. In comparison with other two models, the DTVCN model provides the higher value of average data rate $ x^* $ of users.

In future work, we will provide some real environments by using a network simulator. Network congestion can be studied for varying packet generation rate on different topologies of the networks. 

\bibliographystyle{polonica}
\bibliography{ref_cor}

\begin{thebibliography}{10}

\bibitem{bara}
Albert-L{\'a}szl{\'o} Barab{\'a}si, R{\'e}ka Albert, Hawoong Jeong,
\newblock {\em Physica A: Statistical Mechanics and its Applications}
\newblock {\bf  272}, 173 (1999).

\bibitem{ER}
Paul Erds, Alfr{\'e}d R{\'e}nyi,
\newblock {\em Publ. Math. Inst. Hung. Acad. Sci}
\newblock {\bf  5}, 17 (1960).

\bibitem{TVG}
Klaus Wehmuth, Artur Ziviani, Eric Fleury,
\newblock In {\em Data Science and Advanced Analytics (DSAA), 2015. 36678 2015.
  IEEE International Conference on} IEEE 2015, p.~1.

\bibitem{TVGdynamic}
Arnaud Casteigts, Paola Flocchini, Walter Quattrociocchi, Nicola Santoro,
\newblock {\em International Journal of Parallel, Emergent and Distributed
  Systems}
\newblock {\bf  27}, 387 (2012).

\bibitem{synchronization}
Vivek Kohar, Peng Ji, Anshul Choudhary, Sudeshna Sinha, J{\"u}ergen Kurths,
\newblock {\em Physical Review E}
\newblock {\bf  90}, 022812 (2014).

\bibitem{2007continuum}
Joan Salda{\~n}a,
\newblock {\em ibid.}
\newblock {\bf  75}, 027102 (2007).

\bibitem{2015partial}
Timoteo Carletti, Floriana Gargiulo, Renaud Lambiotte,
\newblock {\em The European Physical Journal B}
\newblock {\bf  88}, 18 (2015).

\bibitem{kumari2016}
Suchi Kumari, Anurag Singh, Priya Ranjan,
\newblock In {\em Proceedings of the International Conference on Internet of
  things and Cloud Computing} ACM 2016, p.~12.

\bibitem{kumari2017}
SUCHI KUMARI, ANURAG SINGH,
\newblock {\em Advances in Complex Systems}
\newblock {\bf  20}, 1850006 (2017).

\bibitem{tee2016}
Philip Tee, Ian Wakeman, George Parisis, Jonathan Dawes, Istv{\'a}n~Z Kiss,
\newblock {\em The European Physical Journal B}
\newblock {\bf  90}, 226 (2017).

\bibitem{BB}
Ginestra Bianconi, Albert-L{\'a}szl{\'o} Barab{\'a}si,
\newblock {\em Physical review letters}
\newblock {\bf  86}, 5632 (2001).

\bibitem{networkscience}
Albert-L{\'a}szl{\'o} Barab{\'a}si,
\newblock {\em Network science},
\newblock Cambridge university press 2016.

\bibitem{multirobust}
Byungjoon Min, Su~Do~Yi, Kyu-Min Lee, K-I Goh,
\newblock {\em Physical Review E}
\newblock {\bf  89}, 042811 (2014).

\bibitem{2016analysis}
Kazuhiro Takemoto, Tatsuya Akutsu,
\newblock {\em PloS one}
\newblock {\bf  11}, e0157868 (2016).

\bibitem{multiplex}
Sen Nie, Xuwen Wang, Binghong Wang,
\newblock {\em Physica A: Statistical Mechanics and its Applications}
\newblock {\bf  436}, 98 (2015).

\bibitem{Holme2016}
Chao-Ran Cai, Zhi-Xi Wu, Michael~ZQ Chen, Petter Holme, Jian-Yue Guan,
\newblock {\em Physical review letters}
\newblock {\bf  116}, 258301 (2016).

\bibitem{newmanassortative}
Mark~EJ Newman,
\newblock {\em ibid.}
\newblock {\bf  89}, 208701 (2002).

\bibitem{ranrewiring}
Jing Qu, Sheng-Jun Wang, Marko Jusup, Zhen Wang,
\newblock {\em Scientific reports}
\newblock {\bf  5}, 15450 (2015).

\bibitem{kelly}
Frank Kelly,
\newblock In {\em Mathematics unlimited—2001 and beyond},
\newblock Springer 2001,
\newblock p. 685.

\bibitem{la}
Richard~J La, Venkat Anantharam,
\newblock {\em IEEE/ACM Transactions on Networking (TON)}
\newblock {\bf  10}, 272 (2002).

\bibitem{incentives}
Asuman Ozdaglar, R~Srikant,
\newblock {\em Algorithmic Game Theory}
\newblock {\bf  647}, 571 (2007).

\bibitem{chen2004}
Qinghua Chen, Dinghua Shi,
\newblock {\em Physica A: Statistical Mechanics and its Applications}
\newblock {\bf  335}, 240 (2004).

\bibitem{kumari2017optimal}
Suchi Kumari, Anurag Singh, Hocine Cherifi,
\newblock In {\em International Computing and Combinatorics Conference}
  Springer 2017, p. 642.

\bibitem{kumari2016modeling}
Suchi Kumari, Anurag Singh,
\newblock In {\em International Workshop on Complex Networks and their
  Applications} Springer 2016, p.~29.

\bibitem{dynamicalcor}
Romualdo Pastor-Satorras, Alexei V{\'a}zquez, Alessandro Vespignani,
\newblock {\em Physical review letters}
\newblock {\bf  87}, 258701 (2001).

\bibitem{onset}
Liang Zhao, Ying-Cheng Lai, Kwangho Park, Nong Ye,
\newblock {\em Physical Review E}
\newblock {\bf  71}, 026125 (2005).

\bibitem{rateeqn}
Roger Guimer{\`a}, Albert D{\'\i}az-Guilera, Fernando Vega-Redondo, Antonio
  Cabrales, Alex Arenas,
\newblock {\em Physical review letters}
\newblock {\bf  89}, 248701 (2002).

\end{thebibliography}
\end{document}